\documentclass[slac_one]{revtex4}
\usepackage{graphicx}
\usepackage{fancyhdr}
\def\etal       {{\it et al}}

\def\ONBB      {{$0\nu\beta\beta$}}
\def\TNBB      {{$2\nu\beta\beta$}}
\def\ONBBD      {{$0\nu\beta\beta$ decay}}
\def\MBB        {{$|m_{\beta\beta}|$}}
\def\gesix      {{$^{76}$Ge}}
\def\sesix      {{$^{76}$Se}}

\def\LN         {LN$_2$}
\def\lar        {LAr}

\def\vctsper    {{$10^{-4}$~counts/(keV$\cdot$kg$\cdot$y)}}

\def\CUORI     {{\mbox{{\sc Cuoricino}}}}

\def\GEM       {{\mbox{{\sc Gem}}}}
\def\GENI      {{\mbox{{\sc Genius}}}}
\def\GERDA     {\mbox{{\sc Gerda}}}
\def\HDM       {\mbox{{\sc HdM}}}
\def\IGEX      {{\mbox{{\sc Igex}}}}
\def\LNGS      {{\mbox{{\sc Lngs}}}}
\def\LVD       {{\mbox{{\sc Lvd}}}}

\def\NEMO      {{\mbox{{\sc Nemo3}}}}

\begin{document}

\title{Search for neutrinoless double-beta decay of Ge-76 with GERDA} 

\author{K.T. Kn\"opfle for the \GERDA\ collaboration}
\affiliation{Max-Planck-Institut f\"ur Kernphysik,  Saupfercheckweg 1, D-69117 Heidelberg, Germany}

\begin{abstract}
 \GERDA , the GERmanium Detector Array experiment, is a 
 new double beta-decay experiment which is currently under construction 
 in the INFN National Gran Sasso Laboratory (\LNGS ), Italy. It is implementing 
 a new shielding concept by operating bare Ge diodes - enriched in Ge-76 - 
 in high purity liquid argon supplemented by a water shield. The aim 
 of \GERDA\ is to verify or refute the recent claim of discovery, 
 and, in a second phase, to achieve a two orders of magnitude lower 
 background index than recent experiments. The paper discusses motivation, 
 physics reach, design and status of construction of \GERDA , and presents 
 some R\&D results.
 \end{abstract}

\maketitle

\thispagestyle{fancy}

\section{INTRODUCTION}

Neutrino physics can help to answer fundamental questions in subatomic physics, astrophysics 
and cosmology. 
The recent discovery of non-zero neutrino masses has provided already first evidence for new 
physics beyond the standard model. However, many neutrino properties remain to be measured 
and even the absolute neutrino masses are still unknown.  
The \GERDA\ experiment \cite{GLOI,GPRO} will search for neutrinoless double-beta (\ONBB ) decay in 
\gesix ,
a  lepton number violating process in which the nucleus \gesix\ of charge Z=32 decays into \sesix\
with charge Z=34 and two electrons. 
It can be viewed as the familiar
\TNBB\ decay, Z $\rightarrow$ (Z+2) + 2\,e$^-$ + 2\,$\bar \nu_e$, where the two 
anti-neutrinos annihilate. The observation of \ONBBD \ would establish the neutrino to be its
own anti-particle, or Majorana particle. 
Its half-life, 
(T$^{0\nu}_{1/2}$)$^{-1}$ = G$_{0\nu}\cdot$$|$M$_{0\nu}|^2\cdot |m_{\beta\beta}|^2$, depends on a
phase space factor G$_{0\nu}$, the nuclear matrix element M$_{0\nu}$, and the effective Majorana mass 
$|m_{\beta\beta}|$ = $|\Sigma_i\,U^2_{ei}m_i|$, where $U_{ei}$ are elements of the neutrino
mixing matrix and $m_i$ the masses of the neutrino mass eigenstates. 
So, a measurement of the \ONBB\ half-life will yield information about the absolute neutrino mass 
scale if the left-hand weak current is the dominant source for \ \ONBBD . 

The experimental signature for \ONBBD\ is the observation of a peak at the endpoint Q$_{\beta\beta}$ in
the energy spectrum of the 2\,e$^-$ final state. Diodes fabricated from high purity Ge (HPGe) material 
enriched in \gesix\ are outstanding $\beta\beta$ detectors
being simultaneously the \ONBBD\ 
source and a 4$\pi$ detector with the excellent energy resolution of a few keV at Q$_{\beta\beta}$ = 2039 keV. 
The best limits for \ONBBD\ in \gesix\ are due to the Heidelberg-Moscow (\HDM ) and \IGEX\ 
enriched \gesix\ experiments \cite{HDM01,IGX02} yielding lower half-life limits of about 
T$^{0\nu}_{1/2} > 1.6\cdot 10^{25}$~y and  corresponding effective Majorana masses of 
\MBB $<$ 0.33\,-\,1.35~eV where the range of \MBB\ values reflects the estimated uncertainties in the 
nuclear matrix elements needed to convert T$^{0\nu}_{1/2}$ into \MBB .
A fraction of the \HDM\ collaboration has claimed recently the observation of \ONBBD\ in \gesix\ with  
a half-life of T$^{0\nu}_{1/2} = 1.2^{+3.0}_{-0.5}\cdot 10^{25}$~y (3$\sigma$ range), implying a \MBB\ value 
between 0.1 and 0.9~eV with the central value of 0.44~eV \cite{Kla04}. In view of the controversial
aspects of this result, see e.g. refs. \cite{Bak05,Str05}, scrutiny by other more sensitive experiments is
needed.  
The ongoing experiments \CUORI\ \cite{Arn05} and \NEMO\ could confirm the \ONBBD\ signal with $^{130}$Te 
and  $^{100}$Mo but cannot refute the claim in case of a null result \cite{Arn05} due to the uncertainties 
of the nuclear matrix elements, see \cite{Rod06} for discussion. 
The \GERDA\ experiment  aims at probing \ONBBD\ of \gesix\ with a sensitivity of 
T$^{0\nu}_{1/2} > 1.4\cdot 10^{26}$~y at 90\% confidence level corresponding to a \MBB\ range from
0.1 to 0.3~eV. Using in its first phase the refurbished \gesix\ detectors of 
the previous \HDM\ and \IGEX\ experiments, a total of about 18~kg, 
\GERDA\ will be able to scrutinize the recent claim for the \ONBBD\ observation with high statistical 
significance after one year of running. \GERDA\ will reach its
ultimate sensitivity in phase II where the total \gesix\ mass will be increased beyond 30~kg 
by adding custom made detectors.

\section{Experimental Setup}

\begin{figure*}[tb]
\centering
\includegraphics[width=150mm]{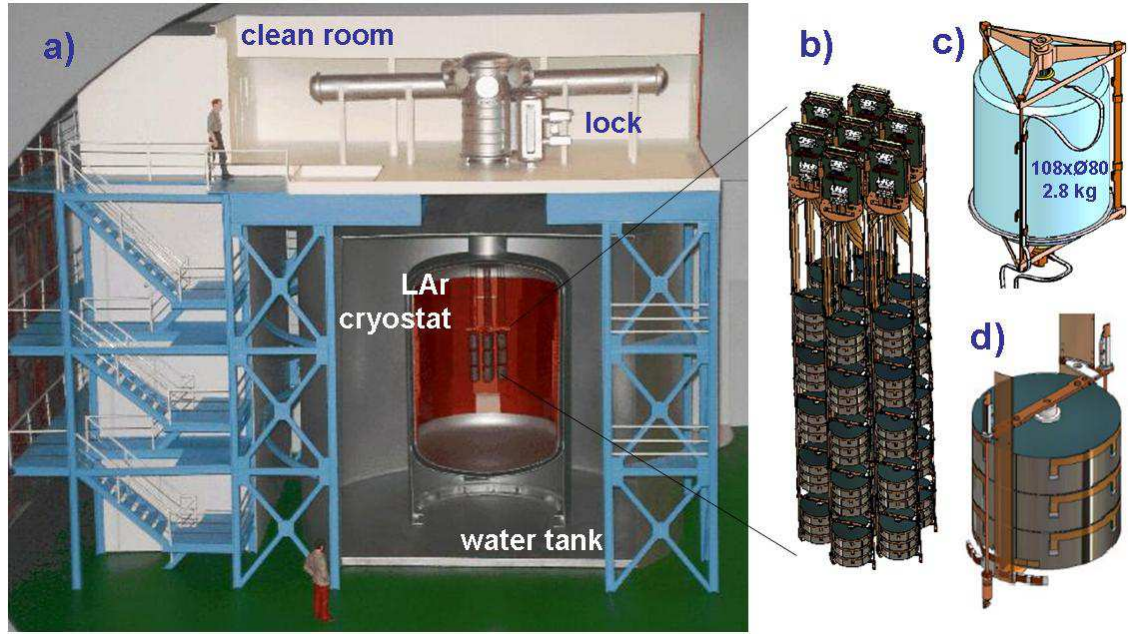}
\caption{Artist's views of
a) the \GERDA\ experiment,
b) a 5$\times$7 Ge detector array assembled from 7 strings of 5 diodes,
c) a p-type Ge diode for phase I and d) a segmented true coaxial n-type Ge diode for phase II in
low mass support and contact structures. Laboratory rooms behind the staircase are allocated for
the water plant and a radon monitor (ground floor), a control room (first floor), and for
vacuum, cryogenic and electronic components (second floor).} \label{gerda1}
\end{figure*}

The \GERDA\ experiment implements an earlier proposal \cite{Heu95} to operate bare Ge diodes
in an ultra-pure liquefied gas, liquid nitrogen (\LN ) or argon (\lar ). The cryogenic
liquid acts not only as cooling medium for the diodes but represents also an unsurpassed shield 
against the {\it external gamma background} that has dominated in previous experiments.
Figure \ref{gerda1}a) shows a schematic of the experimental setup that is being built in 
Hall A of \LNGS , 
below about 3800 meter water equivalent of rock of the Gran Sasso mountain.
A superinsulated cryogenic vessel of  4~m diameter is immersed in a water tank with a diameter
of 10~m and an effective height of 8.5~m. A similar graded shield has been discussed in the
\GEM\ proposal \cite{Zde01}. The 3\,m thick layer of highly purified water reduces the
radioactivity of the rock and concrete  ($\sim$3~Bq/kg $^{228}$Th) well below that of the
cryostat walls which is then reduced by the 2~m thick cryogenic shield to the desired
background index (BI) of a few \vctsper\ \cite{igor}.  
The water buffer serves also as a neutron shield and, instrumented 
with photomultipliers, as Cherenkov detector for efficiently vetoing cosmic muons \cite{Pan07}.
The Ge detector array, Fig.~\ref{gerda1}b), has a hexagonal structure and is made up of individual 
detector strings. 
A detector string is assembled from up to five 
independent Ge detector modules. Designs of such modules are shown in Fig.~\ref{gerda1}c) and d) for 
p-type (phase I) and segmented true coaxial n-type Ge diodes (phase II). 
A cleanroom and lock on top of the vessel assembly allow to insert and 
remove individual detector strings without contaminating the cryogenic volume. Similarly, calibration
sources can be brought close to the array. Radon tightness throughout the experimental volume is 
achieved by the exclusive use of metal seals in all relevant components. 

The {\it intrinsic backgrounds} of the Ge diodes need also to be reduced in order to yield the
desired level of sensitivity.  The most relevant contributions are known to come from $^{60}$Co 
and $^{68}$Ge 
since their lifetimes are in the order of years and their decay chains exhibit Q-values above the 
Q$_{\beta\beta}$ value. Among others, these isotopes are produced by cosmogenic spallation in the 
germanium. 
An obvious way for their reduction is to limit the exposure of the enriched Ge material  
to (hadronic) cosmic rays as much as possible. This recipe is being followed in the procurement of new 
enriched detectors. 
Other methods for intrinsic background suppression will exploit that $\beta\beta$ decay has a 
point-like energy deposition 
while the $^{60}$Co and $^{68}$Ge decays 
lead to  extended events due to multiple Compton interactions. Such 'multi-site' events will be suppressed by 
anti-coincidence of detectors within the array or, due to higher granularity even more efficiently, 
in segmented detectors.   Another complementary method is to identify 
multi-site events from the time structure 
of the signal. With LAr as cryogenic fluid, these backgrounds can additionally be suppressed very 
efficiently by detecting the scintillation light of  LAr.

\section{R\&D ACTIVITIES}

\begin{figure*}[htb]
\centering
\includegraphics[width=145mm]{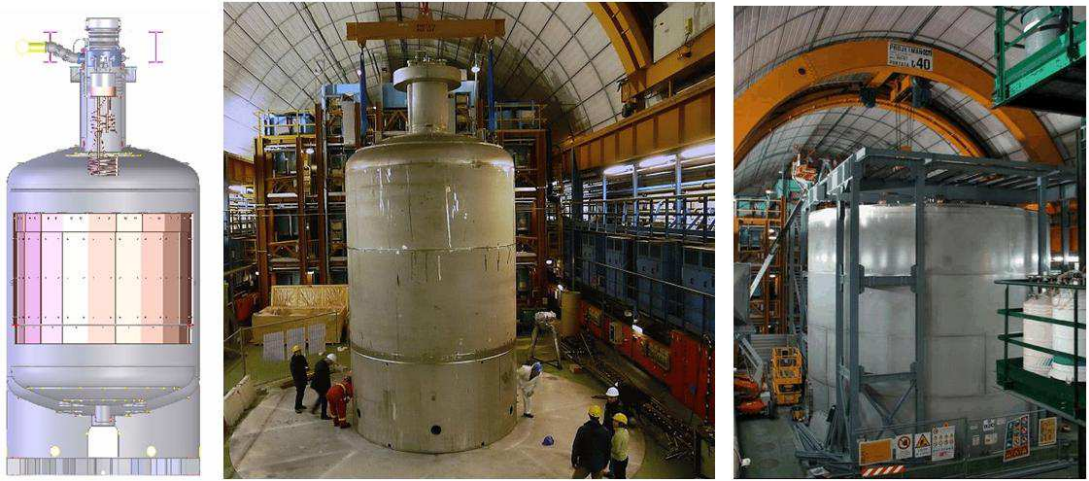}
\caption{ LAr cryostat with copper shield and active cooling system (left), 
the cryostat mounted on the water tank's bottom plate in Hall A (middle), and the 
construction of the \GERDA\ building around the finished water tank which houses the cryostat.}  
\label{gerda2}
\end{figure*}

Major {\it mechanical engineering} has been devoted to the design and clean manufacture of the cryostat
which is realized as a double-walled super-insulated pressure vessel with a nominal volume of 64 m$^3$. 
Originally, the cryostat was planned to be fabricated from special OFE copper 
($<20\mu$Bq/kg $^{228}$Th) by electron beam welding. However, an unexpected increase of cost 
forced the implementation of the backup option, a stainless steel (1.4571) cryostat whose inner 
cylindrical shell is covered by ultrapure OFE copper, see Fig.~\ref{gerda2},\,left). This approach implies the 
use of LAr to limit the mass of the copper shield. A detailed screening of all steel plates
yielded an unexpectedly low radioactivity from less than 1 to 5 mBq/kg $^{228}$Th \cite{maneschg}; hence
the copper shield of only 6 cm thickness includes even a safety margin  \cite{igor}. 
Various acceptance tests of the cryostat include its certification as a pressure vessel according to 
AD2000, the measurement of its thermal loss ($<$300\,W), and the verification that the eight pads
supporting the inner vessel share all the same load.
The Rn emantion rate has been measured to be (14$\pm$2) and (34$\pm$6)\,mBq 
before and after the mount of the copper shield, respectively. Since 10 mBq of Rn, homogeneously
distributed in the cryogenic volume, contribute \vctsper\ to the BI, a final cleaning cycle will be
done. To avoid contamination by frequent refills, the cryostat is equipped with an active cooling system
(Fig.~\ref{gerda2},\,left)).
As the cryostat is operated in a tank filled with water, special safety measures have been taken
including a detailed risk analysis, earth-quake tolerance up to 0.6\,g, the ban of penetrations within the 
cryogenic volume, thermal insulations at inner and outer shells to limit the evaporation rate in case
of leakage of one wall, and, last not least, a nominal design for 1.5 bar overpressure while the operating
pressure will be 0.2 barg.    

A clean underground detector test facility has been installed at \LNGS\ for {\it detector R\&D}.
There all  \HDM\ and \IGEX\ diodes to be used in phase I of \GERDA\ have been characterized 
\cite{oleg}, and refurbishment and handling procedures have been developed and verified. The observed reversible
radiation-induced increase of leakage current in \lar\ can be reduced to an insignificant effect
by reducing the size of the passivation layer.
By now, all enriched diodes as well as those from the \GENI\ test facility
have been successfully refurbished. Moreover, the formerly reported problem of a \lq limited long-term
stability of naked detectors in liquid nitrogen\rq\ \cite{kla06} could not be verified. 
Studies with three non-enriched p-type HPGe diodes during two years of operation with more than 50 
warming and cooling cycles have led to the definition of an optimum detector 
handling procedure. By following this procedure, the detector performance has been proven to be 
stable over the long-term measurements. - 
For phase II detectors, material (37.5 kg)  has been enriched ($>86\,\%$ of \gesix ) in Russia, 
transported in a shielded container to Munich and stored underground in order to minimize the 
cosmogenic production of radionuclides. 
Test purifications of natural GeO$_2$ and its reduction to 6N metal have yielded \cite{ppm} the 
outstanding yield of 90\%. For crystal pulling and characterization a collaboration has been started 
with the Institut f\"ur Kristallz\"uchtung in Berlin, and a first natural Ge crystal has been pulled
with the dedicated puller. -
For phase II, a 3x6 fold segmented  n-type true-coaxial diode of natural Ge has been tested;  
with a novel low mass contacting scheme \cite{Abt07} it exhibits a resolution of 3\,keV at 1.3 MeV for both 
core and segments. The functioning of these contacts has been verified also in \LN . 
Detailed studies have established its power for discriminating single- and multi-site events.
For inside $^{60}$Co and  $^{68}$Ge impurities, the suppression factor is more than an order of 
magnitude \cite{Abt06}. 
As pointed out recently, also a modified electrode HPGe detector can exhibit very good
pulse shape discrimination properties \cite{barb} without any segmentation.
Pursuing this line, first tests have shown that a  standard commercial product, the Canberra 
Broad Energy Ge (BEGe) Detector \cite{bege}, exhibits similar characterisitics and might be indeed a 
cost-effective alternative to a segmented detector.

{\it Electronic engineering} has produced ASIC charge sensitive preamplifiers for readout of the
HPGe diodes at 77\,K. The
PZ-0 circuit is built in the AMS HV 0.8 $\mu$m CZX process and has discrete input FET and feed-back
components. It fulfills all requirements including  a bandwith of 20 to 30 MHz, an equivalent noise 
charge of less than 150e at 30\,pF and a rise time of 15\,ns after a coaxial cable of 10\,m length.
A second fully integrated version is under test.
       
\section{STATUS AND OUTLOOK}

Constituted in February 2004, the \GERDA\ collaboration comprises now 
about 95 physiscists from 17 institutions of six countries. The Letter of Intent 
\cite{GLOI} was submitted to \LNGS\ in March followed by the Proposal \cite{GPRO}
in September 2004. \LNGS\ has approved \GERDA\ in February 2005, allocated space 
for it in Hall A in front of the \LVD\ detector, and acknowledged the safety concept of 
\GERDA . After the delivery of the cryostat in March 2008, the construction of the
\GERDA\ experiment continued with the construction of the water tank and  the \GERDA\
laboratory building including the platform for cleanroom and lock (Fig.~\ref{gerda2}). 
The installation of
these latter parts is expected early in 2009. It will be followed by the commissioning
of the \GERDA\ experiment and the start of phase I data taking.

\end{document}